\documentstyle[12pt]{article}

       \def\s{\sqrt}
        \def\be{\begin{equation}}
        \def\si{\sigma}
        \def\S{\Sigma}
        \def\o{\over}
        \def\b{\bar}
        \def\r{\rho}
        
        \def\c{\chi}
        \def\p{\partial}
        \def\L{\Lambda}
        \def\l{\lambda}
        \def\h{\hat}
        \def\e{\eta}
        \def\m{\mu}
        \def\n{\nu}
        \def\F{\Phi}
        \def\x{\xi}
        \def\D{\Delta}
        \def\i{\int}
        \def\d{\delta}
        \def\g{\gamma}
        \def\ee{\end{equation}}

\begin{document}
\begin{titlepage}
\vspace*{5mm}
\begin{center} {\Large \bf Confinement and screening of the Schwinger model \\
\vskip 0.5cm
on the Poincare half plane}
\vskip 1cm
{\bf H. Mohseni Sadjadi$^a$ \footnote {e-mail:amohseni@khayam.ut.ac.ir},
M. Alimohammadi$^{a,b}$ \footnote {e-mail:alimohmd@theory.ipm.ac.ir}}
\vskip 1cm
{\it $^a$Physics Department, University of Tehran, North Karegar,} \\
{\it Tehran, Iran }\\
{\it $^b$Institute for Studies in Theoretical Physics and Mathematics,}\\
{\it  P.O.Box 5531, Tehran 19395, Iran}

\end{center}
\vskip 2cm
\begin{abstract}
We discuss the confining features of the Schwinger model on the Poincare half
plane. We show that despite the fact that the expectation value of the
large Wilson loop of massless Schwinger model displays the
perimeter behavior, the system can be in confining phase due to
the singularity of the metric at horizontal axis. It is also shown that
in the quenched Schwinger model, the area dependence of the Wilson
loop, in contrast to the flat case, is a not a sign of confinement
and the model has a finite energy even for large  external charges separation.
The presence of dynamical fermions can not modify the screening
or the confining behavior of the system. Finally we show that in the
massive Schwinger model, the system is again in screening phase.
The zero curvature limit of the solutions is also discussed.
\end{abstract}
\end{titlepage}
\newpage

\section{ Introduction }
One of the exactly soluble model in quantum field theory is the quantum
electrodynamics of the massless fermions in 1+1 dimensions, which is
known as Schwinger model [1]. The massive Schwinger model, describing
the electromagnetic interaction of a massive Dirac field, is no longer
exactly soluble, however all non-trivial features of the massless model
continue to hold for small fermion mass limit [2]. The Schwinger model may
serve as a laboratory to study some important features of particle physics,
present also in higher dimensional theories, such as screening and quark
confinement which are some of the most important problems in particle
physics. For example it has been proposed that the infrared behavior
of QCD$_4$ may be responsible for the confinement of quarks and gluons.
But the concept of infrared slavery, i.e. the increase of potential
between colored objects with separation, could not be verified to be true
using the perturbation methods (because of infrared singularities) and
must be studied nonperturbatively. These kinds of calculations can be done
in an equivalent two-dimensional model.

As it is well known, the gauge field
of the massless Schwinger model can be made massive by the standard Higgs
mechanism and the Coulomb force is replaced by a finite range force.
Then by introducing two opposite static external charges $q$
and ${\bar q}$, one can see that the potential tends to some
constant for large separation of $q{\bar q}$ pairs,
reflecting the screening of these charges by the induced
vacuum polarization. On the other hand in the massive Schwinger model, a
semiclassical analysis reveals a linear $q{\bar q}$ potential.
In this case, by computing the Wilson loop for widely separated
charges, within the framework of Euclidean path integral and
mass perturbation (for small masses), one can see that integer
external probe charges are completely screened whereas a linearly
potential is formed between widely separated non--integer charges [3].

A particular intriguing and interesting case occurs when the two dimensional
surface, on which the model is defined, is a curved space-time.
(Similar investigations for pure Yang-Mills theories on arbitrary two
dimensional compact Riemann surfaces have been done in several papers,
see for example [16,20,21].)
These models
are useful for better understanding the confinement and screening mechanisms
in curved space-time and can be viewed as a first step to study these
phenomena in the presence of quantum gravity. Moreover, they may
have application in string theory and quantum gravity
coupled to nonconformal matter (note that the kinetic term of
the gauge field spoils the conformal invariance of the theory).

The Schwinger model has been studied on different non flat
surfaces, for example on closed Riemann surfaces of genus $g\geq 2$ [4],
on torus [5], and on sphere [6]. Also the Green function of the gauge
field of the Schwinger model has been calculated on the Poincare disk in [7].
Moreover, in [22], the authors have considered a $D$--dimensional hyperboloid
with negative curvature, embedded in $(D+1)$--dimensional Minkowski space,
and by considering the behavior of the gauge and matter fields near the
boundary, they have chosen the solutions with suitable behavior.
In this way, they have used the negative curvature
space--time as a regulator for interacting Euclidean quantum field theories.
However, the
confinement and screening properties of the Schwinger model have not yet been
studied on curved space--time. In [8], where the bosonization
procedure of the Schwinger model in curved space has been discussed, it has
been mentioned that this model continues to exhibit screening
or confinement of the charges associated to the electromagnetic
field on conformally flat spaces.
As we will show, it is not true at least for the Poincare half plane.
In [22], the authors have argued that as
the perimeter and the area in hyperboloid space--times are proportional
for large loops, one can not simply distinguish between different phases
by only considering the Wilson loop dependence on area or perimeter. As we will show, this is
right, i.e. by explicit computation of effective static potential
between a quark and antiquark, we show that despite the different
behavior of the Wilson loop of the Schwinger model in $e\neq 0$
and $e=0$ (the first has perimeter behavior, while the second
has area behavior), both have a common phase structure.

In this paper we want to study the confining behavior of the Schwinger model
on the Poincare half plane. This is an interesting case because
it can illustrate the effects of the boundary {\it and} the metric of
the space--time on the confinement feature of the Schwinger model.
Other property of the Poincare half plane is that
its metric is independent of one of the coordinates, so one can obtain
the static potential of the external charges in terms of the
spatial geodesic distance.

The paper is organized as follows. In section 2, following the method
used in [3], we obtain an expression for the potential between
the external charges by integrating out the fermionic degrees of
freedom. We discuss the confining and screening
like behaviors of the system and point out the differences
of these features with respect to the flat case. We justify
our results by calculating the expectation value of the Wilson loop.
We also derive the bosonization rules for the Schwinger
model on the Poincare half plane.
In section 3 we consider the massive Schwinger model.
Using the bosonization method and by solving the equations of motion
of the gauge and matter fields, we obtain a perturbative expression
for the interaction energy of the probe charges.\\
Note that in this paper we do not consider the nontrivial
topologically sectors of the gauge fields.

\section{Massless Schwinger model on the Poincare half plane and
its confining behavior }

The Poincare half plane, $H=\{(x,t),x>0\}$, is a non--compact Riemann surface
equipped with the metric $ds^2=(dx^2+dt^2)r^2/x^2$ and the symplectic area
form $\s {g}d^2x= (dx{\wedge dt})r^2/x^2$. $r$ is a scale parameter of the
Poincare plane and is related to
the scalar curvature by $R= -2/{r^2}$.
This space is conformally related to the compact orientable Riemann surface
$\S_{g}$ with genus $g\geq 2$, $\S_{g}$=H/G, where G is a discrete
subgroup of PSL(2,R)
(the isometry group of H). The geodesics of the Poincare half plane
are semi--circles centered on the horizontal axis $t$ (which we take it as
the time axis), and straight lines parallel to the vertical axis $x$.
The geodesic distance between the points $(x_1, t_1)$
and $(x_2, t_2)$ on the semi circle is
\be
L=r{\rm cosh}^{-1}[1+{{(x_{2}-x_{1})^2+(t_{2}-t_{1})^2}\o {2x_{1}x_{2}}}].
\ee
For the points $(x_1, t)$ and $(x_2, t)$, $x_{2}>x_{1}$,
situated on the straight line the geodesic distance is given by
\be
d=r{\rm ln}{x_{2}\o x_{1}}.
\ee

The Schwinger model is defined by the action
\be
S= \i \s {g}d^2x[-i{\b \psi }{\h \g }^ae^{\m }_a(\p _{\m }-
ieA_{\m })\psi+ {1\o 4}g^{\m \r}g^{\n \l}F_{\m \n}F_{\r \l}],
\ee
where $e$ is the charge of dynamical fermions, and ${{\hat \gamma}^a}$ are
anti--Hermitian matrices which in terms of Pauli matrices are
${\h \g}^0= i\sigma_{2}$ and ${\h \g}^1= i\sigma_{1}$. $F_{\mu \nu }$
is defined by
$F_{\m \n }= \p_{\m}A_{\n}-\p_{\n}A_{\m}$, $g_{\m \n}= \e_{\m \n}r^2/x^2$,
${\e _{\m \n}}={\rm diag}(1,1)$, and $\s {g}= r^2/x^2$.
The zwei-beins fields ($e^{\m}_a, e^a_{\m}$) are defined through
\be
g_{\m \n}= e^a_{\m}e_{\n}^{b}\e_{ab}, \ \ \ \ \ g^{\m \n}
= e^{\m}_{a}e^{\n}_{b}\e^{ab}.
\ee
For the metric $g_{\m \n }= \e_{\m \n}r^2/x^2$, we obtain
$$
e_{\m}^a= {r\o x}{\d^a_{\m}}, \ \ \ \ \ e^{\m}_{a}={x\over r}\d^{\m}_{a},
$$
\be
e^{\m a}={x\over r}\e^{\m a}, \ \ \ \ \ e_{\m a}={r\o x}\e_{\m a}.
\ee
The action (3) is invariant under change of coordinate system, frame rotation
$e^a_{\m}\rightarrow  \L ^a_{b}e^b_{\mu}$, $\L \in$ SO(2),
and local gauge transformation, but it is not conformal invariant since
the Maxwell field theory is conformal invariant only in four dimensions.

\subsection{ Bosonization }

The classical equation of motion of the field $A_{\mu}$ is
$$
{1\o \s{g}}\p_{\n}{\s g}F^{\n \si}= J^{\si}
= {-e}{\b \psi}{\h \g}^be^{\si}_{b}\psi,
$$
which yields
$$
\p _{\si}{\s g}{\b \psi}{\h \g}^be^{\sigma}_{b}\psi =0.
$$
Hence
\be
{\b \psi}{\h \g}^b e_b^{\sigma}\psi = {\alpha}\epsilon^{\si \n}\p_{\n}\F ,
\ee
where $\alpha $ is a constant,
$\epsilon^{\si\n}=  {\h \epsilon}^{\si \n}/{\s g}$
and
${\h \epsilon}^{01}= {\h \epsilon}_{01}= 1$, ${\h \epsilon}^{10}= {-1}$.
This relation is one of the bosonization rules for massless fermions in
a two dimensional (conformally flat) space. In [8], it has been shown
that by performing a fermionic change of variables,
$\psi =\c/g^{1/8}$ and ${\b \psi}={\b \c}/g^{1/8}$,
the bosonization of the fermionic part of the action (3) is realized
in a similar method as in the flat case. On the other hand the bosonization
rules on the half plane, R$^{+}\times$R, is the same as
the complete plane [9].
Therefore on the Poincare half plane the bosonization rules are [10]
$$
-i{\b \psi}\g^\m \p_{\m}\psi = {1\o 2}g^{\m \n}\p _{\m}\F \p_{\n}\F ,
$$
\be
{\b \psi}\g^\m \psi = {i\o{\s \pi}}\epsilon ^{\m \n}\p_{\n}\F ,
\ee
$$
{\b \psi}\psi = {1\o{g^{1/4}}}{\b \c}\c = -{1\o{g^{1/4}}}
\S {\rm cos}(2{\s \pi}\F),
$$
in which ${\gamma ^{\m }}= {\hat \gamma }^ae_a^{\m}$,
and ${\Sigma}$ is a $c$--number which depends on the
normal ordering of the composite operator ${\bar \psi}\psi$.
To determine ${\Sigma}$,
we proceed as [11].

On one hand, the bosonization of the composite operator ${\bar \c }\c $
is the same as in the flat case, that is
${\b \c}\c = -\S N_{\m}{\rm cos}(2{\s \pi}\F )$, where $N_{\mu}$
is the normal ordering with respect to the mass $\m = e/{\s \pi}$. Hence
$$
<{\b \c}\c ({\x_{1}}){\b \c}\c ({\x_{2}})>
= {\S}^2<N_{\m}{\rm cos}[2{\s \pi}\F (\x_{1})]
N_{\m}{\rm cos}[2{\s \pi}\F (\x_{2})]>
$$
\be
={{\S^2 r^2}\o {x_{1}x_{2}}}{\rm cosh}[4\pi D(\xi_{1}, \xi_{2})],
\ee
where $\xi_{1}= (x_{1}, t_{1})$, $\xi_{2}= (x_{2}, t_{2})$, are two points
on the upper half plane, and $D(\xi_{1},\xi_{2})$ is the bosonic
propagator [7]
\be
D(\xi_{1}, \xi_{2})= {1\o{2\pi}}Q_l(1+{{2|\xi_{1}- \xi_{2}|^2}
\o{4x_{1}x_{2}}}),
\ee
computed from the Lagrangian
\be
L={1\o 2}g^{\m \n}{\s g}\p_{\m}\F \p_\n \F + {1\o 2}{\m}^2{\s g}{\F}^2.
\ee
$Q_{l}$ is the Legendre function of the second kind and
$l= (-1+{\s {1+4{\m}^2r^2}})/2$.
The appearance of the metric dependent term $r^2/(x_1x_2)$ in eq.(8)
is related to the renormalization of vertex operators on the
curved space--time [12]. In the limit $\xi_{1}\rightarrow \xi_{2}$, we use
the asymptotic behavior of $Q_{l}$ and obtain
\be
<{\b \c}\c ({\xi_{1}}){\b \c}\c ({\xi_{2}})>= {\S^2 r^2\o{2x_{1}x_{2}}}
{\rm exp}[-{\rm ln}{{|\xi_{1}-\xi_{2}|^2}
\o{4x_{1}x_{2}}}-2\g -2\Psi (l+1)],
\ee
where $\gamma $ is the Euler constant, and $\Psi $ is the digamma function.

On the other hand, in the limit $\xi_{1}\rightarrow \xi_{2}$, we have [7]
\be
<{\b \c}\c {(\xi_{1})}{\b \c}\c {(\xi_{2})}>
= {1\o 2{\pi}^2 {|\xi_{1}-\xi_{2}|^2}}.
\ee
Note that this relation is the same as one in flat space--time. The reason of
this equality lies in the fact that in the limit $\xi_{1}\rightarrow \xi_{2}$,
all the $A_{\m}$--dependent terms in evaluating
$<{\bar \c}\c ({\xi_{1}}){\bar \c}\c ({\xi_{2}})>$ are canceled out
[5], and this calculation reduces to one in the free fermion
model, i.e. without gauge field, on a flat Euclidean space--time,
described by the action [8]
\be
S_{\rm free}=\int d^2x(-i{\b \c}{\hat \gamma}^a\partial_a\c ).
\ee
Comparing (11) and (12) we obtain
\be
\S = {1\o {2\pi r}}{\rm exp}[\g +\Psi (l+1)],
\ee
which differs from the result obtained for the complete flat plane:
$(e/2\pi^{3/2}){\rm exp}(\g)$ [10]. This difference is due to the presence
of the curvature which modify the Green function of the
gauge fields appeared in the fermionic
two--point functions [5, 7]. In the limit $R\rightarrow 0$ ($r\rightarrow \infty$),
using $\lim_{x\rightarrow \infty}{\Psi (x)}=\rm{ln}(x)$, we obtain the
same $\Sigma $ as the flat case.

\subsection {Confinement: the effective action approach}

In order to investigate the confining behavior of the action (3), we will
obtain the equation of motion of the gauge field derived from the
corresponding effective action. Using (7), the bosonic version of (3) is
\be
S= \i{\s g}d^2x({{1\o 2}g^{\m \n}\p_{\m}\F \p_{\n}\F - {ie\o \s {\pi}}
{\epsilon^{\m \n}}A_{\m}\p_{\n}\F +{1\o 4}F^{\m \n}F_{\m \n}}).
\ee
Integrating over the bosonic degrees of freedom we arrive at
\be
S_{{\rm eff}}= \int (-{e^2\o 2\pi}{\s g}{F\o {\s g}}{1\o \D}{F\o {\s g}}
+ {1\o{2\s g}}F^2)d^2x,
\ee
in which $\D = (1/\s{g})\p_{\m}g^{\m \n}\s {g}\p_{\n}$, and $F={\h \epsilon}^
{\m \n}\p_{\m}A_{\n}$.

As an alternative method, this effective action can be also
obtained by integrating out the fermionic degrees of freedom of the action (3).
To do this, we should compute the determinant of the Dirac
operator $i\gamma^{\m} D_{\m}= i\g^\m (\p_{\m}-ieA_{\m})$,
\be
D:= -{\rm ln}{{\i D{\b \psi} D\psi{\rm exp}(\i{\s g}d^2x{\b \psi}i\g^\m(\p_\m
-ieA_\m)\psi)}
\o {\i D{\bar \psi} D\psi{\rm exp}(\i{\s g}d^2x{\b \psi}i\g^\m\p_\m\psi)}}
= -{\rm ln}{{{\rm det} i\gamma^\m D_\m}\o {{\rm det}i\gamma^\m\p_\m}}.
\ee
Using the one--loop radiative correction of the two--point function
of the gauge field and also by considering the requirement of the
invariance of the theory under PSL(2,R), it can be shown that [7]
\be
D= -{e^2\o 2\pi}\i{\s g}d^2x{F\o {\s g}}{1\o \D}{F\o \s{g}}.
\ee
Adding the kinetic term of the gauge field, we arrive at (16).

In the gauge $A_1=0$ and in the static case $dA_0/dt=0$, the effective
Lagrangian (16) becomes
\be
{\cal L}_{{\rm eff}}= {e^2\o 2\pi}{A_{0}}^2+ {1\o 2}{x^2\over r^2}({dA_{0}\o dx})^2.
\ee
The above effective Lagrangian density shows that the photon gains a mass
equal to $e/{\sqrt\pi}$, which can be interpreted as a peculiar two--dimensional
version of the Higgs phenomenon.

Now following [3], if we introduce a static external charge
distribution composed of a quark and an anti--quark with charges
$e_{1}= -e'$ and $e_2= e'$ at points $\xi_1=(a,t)$ and $\xi_2=(b,t)$,
respectively, this Lagrangian becomes
\be
{\cal L}= {e^2\o{2\pi}}A_{0}^2+ {1\o 2}{x^2\over r^2}({dA_{0}\o dx})^2
+ J^{0}A_{0}{r^2\o {x^2}},
\ee
where
$$
J^{0}= {i\o{\s{g}}}\sum_{n=1}^2 e_{n}\i \d^2({\xi -\xi_{n}})dt_{n}
$$
\be
= i{x^2\over r^2}e'[\d (x-b)-\d (x-a)].
\ee
In this case, the equation of motion of the field $A_0$ is
\be
{d\o dx}{x^2\over r^2}{dA_{0}\o dx}- {e^2\o \pi}A_{0}= ie'[\d (x-b)-\d (x-a)].
\ee
To find $A_{0}(x)$, we note that the Green function of the self adjoint
operator $P= {d\over dx}{x^2\over r^2}{d\over dx}- {e^2\over \pi}$ is
\be
G_{P}(x,x')= -{r^2\o {2l+1}}{x_{<}^l\o x_{>}^{l+1}},
\ee
where $x_<$ ($x_>$) is the smaller (larger) value of $x$ and $x'$.
This Green function is the same as the Green function of the
radial part of Poisson operator in spherical coordinates
and satisfies the Dirichlet boundary condition at $x=0$
(the Poincare half plane has no boundary, but by boundary condition
we mean the behavior of the fields near the horizontal axis).
In the flat case limit, $r\rightarrow \infty$, $l$ leads to $\mu r$ and
therefore the eq.(23) reduces to \be
-{(g(x)g(x'))^{1/8}\over 2\m }e^{-\m d(xx')},
\ee
where $d$ is the geodesic distance (2). By setting
$g_{\m \n}=\eta_{\m \n}$, eq.(24) leads to the Green function of the
flat case, i.e. $-(1/2\mu)e^{-\mu d}$. Using (23), we obtain
\be
A_{0}(x)= ie'[G_{P}(x-b)-G_{P}(x-a)]
= \cases { -{ie'r^2\o {2l+1}}({b^l\o x^{l+1}}
- {a^l\o x^{l+1}}), &$b<x$  \cr
-{ie'r^2\o {2l+1}}({x^l\o b^{l+1}}- {a^l\o x^{l+1}}), &$a<x<b$  \cr
-{ie'r^2\o {2l+1}}({x^l\o b^{l+1}}- {x^l\o a^{l+1}}), &$x<a.$  \cr }
\ee
To calculate the quark--antiquark energy, we must note that the Schwinger model
on the Poincare half plane can be considered as the analytical continuation
of the corresponding model on a Minkowskian space--time described by the
metric $ds^2=(r^2/x^2)(dt^2 - dx^2)$. By ignoring the $i$ factors in
eqs.(21) and (25) and substituting them back into ${\cal L}^{\rm Min.}$, which
has
the  same form as (20), one can obtain the static external charges energy as
$U= -\int {\cal L}^{\rm Mink.}dx=\int {\cal L}^{\rm Eucl.}dx$ [3]. In this way
we find the interaction energy of the external charges as
\be
U= {1\o 2}\int J^{0}A_{0}{r^2\o {x^2}}dx= {e'^2\o{2l+1}}
{r^2\o 2a}(-2e^{-{d\o r}(l+1)} + e^{-{d\o r}}+1).
\ee
For a detailed discussion on the relation of the Euclidean action with the
Minkowskian static energy, see [17].

In the flat case, the eq.(22) is replaced by
\be
{d^2\over dx^2}A_{0} - {e^2\over \pi}A_{0}= ie'[\delta (x-b) -\delta (x-a)],
\ee
which is invariant under translation ($x\rightarrow {x+c}, c\in$ R),
hence the potential is only a function of charge separation, which is a
translational invariant quantity. But in our case, (22) is not invariant
under scale transformation
(dilatation $x\rightarrow \lambda x$; $\lambda \in$ R),
which leaves the distance $d$ invariant, hence the potential
depends on both the distance
$d=r{\rm ln}(b/a)$ {\it and} the position of the external charges.

For large separation, $b>>a$, the potential tends to
\be
\lim_{d\rightarrow \infty}U= {r^2\o 2a}{e'^2\o {2l+1}},
\ee
which indicates the {\it screening} like phenomenon: By fixing the position
of one of the charges at an arbitrary point $x=a$, and moving the
other charge, the potential increases linearly for small separation
and tends to a finite value for large $d$.
But the crucial point is that the geometry of the Poincare half plane
is non--trivial, and a model defined in this space--time,
may have different behaviors in different
regions. For example, while the confinig phase is dominant in a region,
the system may be in screening phase in another region. To see this,
one must study the behavior of some external charge in this
space, as a probe. Now as it is clear from (26), the system is in confining
phase near the boundary $x=0$:
for $a\simeq 0$,
we must have $b= a+ O(a^2)$ in order to have a finite energy for the system,
otherwise $U\rightarrow \infty$.
This means that in the massless Schwinger model on the Poincare
half plane, the {\it confining phase}, is dominant {\it near the horizontal axis}.
This is related to the singularity of the metric at $x= 0$.

On the flat plane, the Schwinger model is confining in the absence
of dynamical fermions: The screening potential [3]
\be
U_{{\rm flat}}={e'^2\o {2\m}}(1-e^{-\m |b-a|}),
\ee
in the limit $\m \rightarrow 0$, becomes $(e'^2/2)|b-a|$ which increases
with the relative distance of the charges. In this limit, the effects
of the fermionic vacuum polarization is switched off: the screening is
replaced by the confining behavior of the system.

But on the Poincare half plane at $\m =0$, that is when dynamical
massless fermions are absent, the potential becomes
\be
U= {e'^2r^2\o 2a}(1-e^{-{d\o r}}),
\ee
which has the same confining or screening nature as (26). The dynamical
fermions can only decrease the amount of saturated energy.
Hence the screening like (or confining) behavior of the Schwinger model
depends on the vacuum polarization {\it and} the curvature of the space--time.

\subsection {Confinement: the Wilson loop approach }

Now it is interesting to obtain and interpret these results by computing
the Wilson loop expectation value.
The interaction of an external current density $j^\mu$ and the gauge field
$A_{\mu}$ is described by the action
$S_{{\rm int.}}=\int{\sqrt g}j^\mu A_{\mu}d^2x$. We assume that
$j^{\mu}$ is produced by two external charges moving on a loop which is
obtained as follows. Two charges $e'$ and $-e'$ are created at the point
$(x,t)$ and move apart in (Euclidean) time $\tau$
to points $(a,t+\tau)$ and $(b,t+\tau)$.
Then they stay static at their positions for a period of time $T$,
and after that come together to annihilate. In the limit $T>>\tau$,
in which we are interested, this Wilson loop becomes a
rectangle $c$
characterized by $a,b$, and $T$,
on the Poincare half plane. The reason for choosing this kind of Wilson loop
is that in the large $T$--limit, the expectation value of this Wilson loop
becomes
proportional to exp[$-U(d)T$] ($U(d)$ is the static external charge potential)
for time--independent metrics [18]. (See also [19] for the same
calculations on a curved space--time.)
The interaction term of this process is $S_{{\rm int}}= ie' \oint_{c}dx^\mu
A_{\mu}$, and the expectation value of the corresponding Wilson loop is

${<W_{c}[A]>}=$
 \be
{{\i DA_{\alpha}D\F \d (H[A_{\alpha}]){\rm det}[{\d H[A^{\lambda}]
\o{\d \l}}]
{\rm exp}(ie'\oint_{c} A_{\m}dx^{\m}){\rm exp}
[\i (-{1\o 2}(\p_{\m}\F )^2
-{ie\o {\s \pi}}
F\F -{1\o 2}{x^2\o r^2}F^2)d^2x}]\o {\i DA_{\alpha}D\F \d (H[A_{\alpha}])
{\rm det}[{\d H[A^{\l}]\o {\d \l}}]
{\rm exp}[\i (-{1\o 2}(\p_{\m}\F )^2-
{ie\o{\s \pi}}F\F- {1\o 2}{x^2\o{r^2}}F^2)d^2x}]}.
\ee
$H[A_\alpha ]=0$ is the gauge--fixing condition and $\lambda $
parameterizes the gauge transformation $A_\alpha^{\lambda}= A_{\alpha}
+ \partial _{\alpha }\lambda$.
One can show that by using the change of variables $A\rightarrow (F,\eta)$,
$\eta := H[A]$, the Jacobian of this transformation:
\be
DA_{\alpha}= {\rm det}^{-1}[{\delta H[A^{\l}]\o{\d \l}}]
D\eta DF,
\ee
cancels precisely against the ghost determinant [16]. Thus
\be
<W_{c}[A]>=
{{\i DFD\F {\rm exp}[\i (-{1\o 2}(\p_{\m}\F )^2
-{1\o 2}{x^2\o r^2}F^2-{ie\o \s{\pi}}F\F )
d^2x]{\rm exp}(ie'\oint_{c}A_{\m}dx^{\m})}\o
{{\i DFD\F {\rm exp}[\i (-{1\o 2}(\p_{\m}\F )^2-
{1\o 2}{x^2\o r^2}F^2-{ie\o \s{\pi}}F\F )d^2x}]}}.
\ee
Using the Stokes theorem
\be
\oint_{c} A_{\mu} dx^{\mu}= \int_{D}\eta ({\xi})F({\xi})d^2x,
\ee
where $c= \partial D$, and $\e ({\xi})= \cases { 1,&$\xi \in D$  \cr
0,&$\xi \notin D.$  \cr }$, we arrive at
\be
<W_c[A]>= {\rm exp}[-{e'^2\over 2}\int {r^2\over x^2}\eta^2({\xi})d^2\xi
- {e^2e'^2\over 2\pi }\int \eta({\xi})G_{W}({\xi,\xi'})
\eta({\xi'})d^2{\xi}d^2{\xi'}],
\ee
in which the Green function $G_{W}({\xi,\xi'})$ satisfies
\be
[{x^2\o r^2}({d^2\o {dx^2}}+{d^2\o {dt^2}}){x^2\o r^2}- {\m^2}{x^2\o r^2}]G_{W}({\xi ,\xi'})
= \d^2({\xi-\xi'}).
\ee
If we insert the Fourier expansion
\be
G_{W}({\xi,\xi'})={1\o {2\pi}}\i^{\infty}_{-\infty }f_{k}(x,x')e^{ik(t-t')}dk,
\ee
in eq.(36), the coefficients are found as following
\be
f_{k}(x,x')=-{r^4}(xx')^{-3/2}I_{l+{1\o 2}}(kx_{<})K_{l+{1\o 2}}
(kx_{>}).
\ee
$I_{l+{1\o 2}}$ and $K_{l+{1\o 2}}$ are modified Bessel
functions of the first and second kind, respectively.
Performing the integration over $k$ we obtain
\be
G_W({\xi ,\xi'})= {r^4\o {2\pi (xx')^2}}Q_{l}
[{{x^2+x'^2+(t-t')^2}\o {2xx'}}],
\ee
Now as the functional integral in the massless Schwinger model is
Gaussian, and the higher order correlators factorize into product of pair
correlators $<F({\xi})F({\xi'})>$, we have [3]
\be
<{\rm exp}(ie'\oint A_{\m }dx^{\m })>
={\rm exp} [-{e'^2\o 2}{\i \i}_{D}d^2\xi d^2\xi'<F({\xi})F({\xi'})>].
\ee
Comparing (40) with (35) and (39), results
\be
<F({\xi})F({\xi'})>= \d^2 ({\xi-\xi'}){r^2\o {x^2}}-
{{\m^2}r^4\o {2\pi (xx')^2}}Q_{l}({\rm cosh}{L\o r}),
\ee
where $L$ is the geodesic distance between $\xi$ and $\xi'$ (see eq.(1)).
By considering the behavior of $I_{l+{1\o 2}}(x)$ at $x\simeq 0$
($Q_l({\rm cosh} {L\o r})$ at $L= \infty$), one can easily check that the
behavior of $f_k(x,x')$ ($<F({\xi})F({\xi'})>$) is consistent with
the Dirichlet boundary condition imposed on eq.(23).

In the flat case limit, one can show that
$\lim_{r\rightarrow \infty}Q_l[{\rm cosh} (L/r)]= K_{0}(\m L)$, and by setting
$g_{\m\n}\rightarrow \eta_{\m\n}$, the eq.(41)
becomes \be
<F({\xi})F({\xi'})>_{{\rm flat}}=\delta^2({\xi-\xi'})
- {\mu^2\over {2\pi}}K_0(\mu |\xi-\xi'|),
\ee
which is the strength fields correlator on the flat space--time [13].
In the absence of dynamical fermions ($\mu =0$), eq.(42)
reduces to $\delta^2({\xi-\xi'})$
and one obtains the area law for the Wilson loop, which is a characteristic of
a confining potential. In the presence of dynamical fermions,
since $K_0(\mu |\xi-\xi'|)$ decays exponentially as
$e^{-\mu |\xi-\xi'|}$, the correlator exhibits the finite correlation
length (related physically to the screening effect), and the perimeter
law is arisen for large contour [14].
On the Poincare half plane, $K_0$ is replaced by $Q_l({\rm cosh}{L\o r})$
which decays as ${[{\rm cosh}(L/r)]}^{-l-1}$ for large $L/r$. Hence in this
case we have also a finite correlation
length for $<F({\xi})F({\xi'})>$ and, as we will show, the Wilson
loop is perimeter dependent. In fact the area term arising from
the delta function is canceled out by the corresponding term in
the integration of $Q_l$.

Using
\be
\lim_{T\rightarrow \infty} {1\over 2\pi T}\int^T_{0}
e^{ikt}dt\int^T_{0} e^{-ik{t'}dt'} = \delta (k),
\ee
and
\be
f_{0}(x,x')=- {r^4\o {2l+1}}{x_{<}^{l-1}\o {x_{>}^{l+2}}},
\ee
one can obtain the following expression
\be
U=\lim_{T\rightarrow \infty} (-{1\o T}{\rm ln}W)=
{r^2\o {2l+1}}{e'^2\o 2}[{1\o a}+{1\o b}-2({a^l\o b^{l+1}})],
\ee
for the static potential between external charges which is equal to
one obtained in eq.(26).
After some calculations, one can show that the Wilson loop for $T>>b>>a$
is
\be
<W_{c}[A]>= {\rm exp}[-{r^2\over {2l+1}}{{e'^2}\over 2} T({1\over a}
+ {1\over b})+O({1\o {T^\lambda }})],
\ee
where using the hypergeometric representation of the Legendre function,
$\lambda $ is found to be $\lambda = 2l+1$.

But the perimeter of the large Wilson loop $T>>b,a$ is
\be
\oint_{c}{r{\s {dx^2+dy^2}}\o x}= T({r\o a}+{r\o b}),
\ee
therefore eq.(46) shows that for a large contour, the perimeter law
is {\it satisfied}, which is the same behavior as the flat space--time case.

In the quenched Schwinger model ($\m =0$), the Wilson loop expectation value is
\be
<W_{c}[A]>= {\rm exp}[-{{e'^2}\over 2}\int\eta^2 {r^2\over x^2}dxdt]=
{\rm exp}[-{{e'^2}\over 2} Tr^2({1\over a} -{1\over b})].
\ee
But we note that $Tr^2(1/a-1/b)$ is nothing but the area of the
rectangle bounded by the Wilson loop. Therefore on the Poincare half plane,
the Wilson loop of the Schwinger model in the absence of dynamical fermions
(that is the pure QED$_2$) {\it is} equal to the exponential of the area,
like the flat case. But
in contrast to the flat case, the area is not proportional to the geodesic
distance of the charges, and we {\it can not conclude} that the system is in
confining phase. This result is consistent with our discussion after eq.(30).

\section{ Confining aspect of massive Schwinger model on
the Poincare half plane }

The massive Schwinger model, i.e. U(1) gauge theory with massive dynamical
fermions of charge $e$ and mass $m$, is defined by the action
\be
S= \int \sqrt{g} d^2x
[-i{\bar \psi} {\hat \gamma}^a e^\mu_{a}(\partial_{\mu } - ieA_{\mu})\psi
+ m{\bar \psi}\psi+ {1\o 4}g^{\m \n}g^{\r \sigma}F_{\m \r}F_{\n \sigma}].
\ee
This model is not soluble even in the flat case, but in the limit $m<<e$,
the physical quantities may be evaluated using perturbative expansion in
fermions mass.

In conformally flat curved space--time, the bosonic form of
the action (49) is [8]
\be
S= \int \sqrt {g}d^2x[{1\over 2}g^{\mu \nu}{\partial_\mu}\Phi {\partial_\nu}
\Phi
- {ie\over \sqrt \pi}\epsilon^{\mu \nu}A_{\m}{\partial_{\nu}}\Phi - {m\Sigma
\over g^{1\o 4}}{\rm cos}(2\sqrt {\pi}\Phi)
+ {1\over 4} g^{\mu \nu}g^{\r \sigma}
F_{\mu \r}F_{\nu \sigma}],
\ee
where the constant $\Sigma$ is
given by (14). In fact this model is a Sine-Gordon model whose interaction
is position dependent [8].
The confining behavior of this system can be analyzed perturbatively
by expanding the mass term in a power of $\Phi $ [15]. The equations
of motion
followed from the bosonized action (50), in the presence of
external charges (21), are
$$
{d\o dx}{x^2\o r^2}{d\o dx}A_{0}+ {ie\o \s{\pi}}{d\F \o {dx}}= ie'[\d (x-b)-\d (x-a)],
$$
\be
-{d^2\F \o {dx^2}}+ {ie\o \s {\pi }}{dA_{0}\o dx}
+ {2\s {\pi }m\S r\o x}{\rm sin}(2\s {\pi}
\F)= 0,
\ee
where we have used, as before, the Coulomb gauge $A_1=0$ and ${dA_0}/dt=0$. Using
the approximation
${\rm sin}(2{\sqrt \pi}\Phi){\simeq  2\sqrt {\pi }\Phi}$,
and assuming that the field $\Phi $
is a slowly varying field [15], we arrive at
\be
{d\o dx}({x^2\o r^2}+{e^2\o 4{\pi}^2m\S r}x){dA_{0}\o {dx}}
= ie'[\d (x-b)- \d (x-a)],
\ee
$$
{{4\pi m\S r}\o x}\F + {ie\o \s {\pi }}{dA_{0}\o dx}= 0,
$$
with solutions
\be
A_{0}(x)= \cases {0, &$x>b$   \cr
-i{4{\pi}^2m\S e'r\o {e^2}}[{\rm ln}(1+{e^2\o {4{\pi}^2m\S }}{r\o b}) -
{\rm ln}(1+{e^2\o {4{\pi}^2m\S }}{r\o x})], &$a<x<b$   \cr
-i{4{\pi}^2m\S e'r\o {e^2}}[{\rm ln}(1+{e^2\o {4{\pi}^2m\S }}{r\o b}) -
{\rm ln}(1+{e^2\o {4\pi^2m\S }}{r\o a})],  &$x<a$,  \cr }
\ee
and
\be
\F (x)= \cases {0, &$x>b$  \cr
-({e'\o e}){e^2\o {4{\pi}^{3\o 2}m\S}}
({1\o {x\o r}+{e^2\o {4\pi^2m\S}}}) ,&$a<x<b$  \cr
0, &$x<a$.  \cr }
\ee
Therefore we find the potential $U= {1\over 2}\int \r A_{0}dx$ as
\be
U=2{\pi }^2m\S ({e'\o e})^2 r[{\rm ln}(1+{e^2\o {4\pi^2m\S }}{r\o a})
-{\rm ln}(1+{e^2\o
{4{\pi}^2m\S }}{r\o b})].
\ee
By fixing $a$ and increasing the separation of the
charges, $U$ increases and finally achieves the limiting value
$2\pi ^2 m\Sigma (e'/e)^2r{\rm ln}(1+{e^2r\over {4\pi^2 m\Sigma a}})$,
which shows that the system is in the screening phase.
When one of the charges is located near the horizontal axis,
$a\rightarrow 0$, the eq.(55) goes to infinity, unless
$b\rightarrow a$. So the system is in confining phase at $x\simeq 0$, as we expect.
On the other hand, in order to satisfy the conditions that we have assumed
for the field $\Phi $ (to be small and small varying), we must take
\be
({x\o r}+{e^2\over {4\pi^2m\Sigma }})>> ({e'\over e}){e^2\over {4\pi^2m\Sigma }}.
\ee
By expanding $U$ in terms of $r/a$ and $r/b$, we obtain
\be
U= {{e'^2}\o 2}{r^2}
({1\o a}-{1\o b})+ O({r^2\o a^2},{r^2\o b^2}).
\ee
For large $a/r$ and $b/r$, $U$ is proportional
to the area of the Wilson loop characterized by $a, b,$ and $T$. In the flat case,
this behavior is interpreted as a sign of confinement but, as we have discussed
earlier, this is not true for the Poincare half plane.

Finally if we consider the small fermion mass limit $m<<e$, and also $e'<<e$,
the eq.(55) reduces to
\be
U= 2{\pi^2}m\Sigma ({e'\over e})^2r{\rm ln}({b\over a}),
\ee
which is comparable with the corresponding result in the flat case, after
substituting $r$ln$(b/a)\rightarrow (b-a)$
and $\Sigma$(Poincare) $\rightarrow \Sigma$ (flat) [3,15].
Note that the potential (58) is proportional to the geodesic distance
$d=r$ln $(b/a)$, but this is not a sign of confinement as in the flat case.
To see this, note that if one fixes the position of the first charge at
$x=a$ and moves the other charge to a large distance ($b/r\rightarrow
\infty$), the eq.(55) reduces to (for $m<<e$, $e'<<e$)
\be
U ({b\o r}\rightarrow \infty)= {e'^2}{r^2}{1\o 2a},
\ee
which has a finite value, and the system is again in the screening
(and {\it not} confining) phase.

\section{Conclusion}
Let us summarize the main results of the paper:

1- In $m=0$, the Schwinger model on flat space--time is in screening phase,
but on the Poincare half plane, the system is in confining phase
in $x\simeq 0$ region, and in screening phase in
regions far enough
from the horizontal axis ( after eq.(28)),

2- In $m=0$ and $e=0$, the model is confining in flat case (after eq.(29)),
but on the Poincare half plane, the phase depends on the region under study
(after eq.(30)).

3- In $m=0$, the Wilson loop obeys the perimeter (area)--law for $e\neq 0$
($=0$) in both the flat (after eq.(42)) and the Poincare (eqs.(47) and
(48)) cases. But on the Poincare half plane, the area dependence does not indicate
the confining phase (in contrast to the flat case) (after eq.(48)).

4- In $m\neq 0$, the Schwinger model on the flat space--time is in
confining phase but in the Poincare case, the model is in screening phase
(after eqs.(55) and (59)).

\vskip 1cm

\noindent{\bf Acknowledgement}
We would like to thank A. Arfaei for useful discussion and
the research council of the
University of Tehran for partial financial support.


\vskip 1cm
\begin{thebibliography}{99}
\bibitem{1} J. Schwinger, Phys. Rev. {\bf 125} (1962) 397; {\bf 128} (1962) 2425.
\bibitem{2} C. Adam, Phys. Lett. {\bf B363} (1995) 79.
\bibitem{3} D. J. Gross, I. R. Klebanov, A. V. Matytsin, and A. V. Smilga,
Nucl. Phys. {\bf B461} (1996) 109.
\bibitem{4} F. Ferrari, Nucl. Phys. {\bf B439} (1995) 692.
\bibitem{5} I. Sachs and A. Wipf, Ann. Phys. {\bf 249} (1996) 380.
\bibitem{6} A. Basseto and L. Griguolo, Nucl. Phys. {\bf B439} (1995) 327.
\bibitem{7} F. Ferrari, Int. J. Mod. Phys. {\bf A11} (1996) 5389.
\bibitem{8} O. J. P. Eboli, Phys. Rev. {\bf D36} (1987) 2408.
\bibitem{9} A. Liguori and M. Mintchev, Nucl. Phys. {\bf B522} (1998) 345.
\bibitem{10}S. Coleman, Ann. Phys. (N.Y.) {\bf 101} (1976) 239,
S. Coleman, R. Jackiw, and L. Susskind, Ann. Phys. (N.Y.) {\bf 93} (1975) 267.
\bibitem{11} A. V. Smilga, Phys. Lett. {\bf B278} (1992) 371, Phys. Rev.
{\bf D55} (1997) 2443.
\bibitem{12} E. Verlinde and H.Verlinde, Nucl. Phys. {\bf B288} (1987) 357.
\bibitem{13} C. Jayewardena, Helv. Phys. Acta {\bf 61} (1988) 636.
\bibitem{14} A. V. Smilga, Phys. Rev. {\bf D46} (1992) 5598.
\bibitem{15} A. Armoni, Y. Frishman, and J. Sonnenschein, Int. J. Mod. Phys.
{\bf A14} (1999) 2475.
\bibitem{16} G. Thompson , " Topological gauge theory and Yang-Mills theory "
, in Proceeding of the 1992 Trieste Summer School on High Energy Physics and
Cosmology , World Scientific , Singapore (1993) , 1-76.
\bibitem{17} R. Rajaraman, "Solitons and instantons, an introduction to
solitons and instantons in quantum field theory", Elsevier, North--Holand,
1989.
\bibitem{18} M. Le Bellac, "Quantum and statistical field theory", Clarendon
Press, Oxford, 1991.
\bibitem{19} J. Sonnenschein, {\it What does the string/gauge correspondence
teach us about Wilson loops?}, hep--th/0003032.
\bibitem{19} E. Witten, Commun. Math. Phys. {\bf 141} (1991) 153.
\bibitem{20} M. Alimohammadi and M. Khorrami, Int. J. Mod. Phys. {\bf A12} (1997)
1959.
\bibitem{22} C. G. Callan and F. Wilczek, Nucl. Phys. {\bf B340} (1990) 366.
\end{thebibliography}
\end{document}